 \documentstyle[prd,eqsecnum,preprint,tighten,aps]{revtex}
 \renewcommand{\theequation}{\thesection.\arabic{equation}}
 \def\appendixa{
 \vskip 1cm
 {\bf APPENDIX A: Useful Algebraic Relations}
 \vskip 1cm
 \par
 \setcounter{equation}{0}
 \def\theequation{A.\arabic{equation}}
 }
 \def\appendixb{
 \vskip 1cm
 {\bf APPENDIX B: The Rindler Space-Time}
 \vskip 1cm
 \par
 \setcounter{equation}{0}
 \def\theequation{B.\arabic{equation}}
 }
 
\begin{document}

 \title{Entangled Quantum Fields near the Event Horizon, \\
 and Entropy}
 \author{A. Iorio$^{a,}$\thanks{E-mail: iorio@lns.mit.edu},
 G. Lambiase$^{b,c,}$\thanks{E-mail: lambiase@sa.infn.it},
 G.Vitiello$^{b,c,d,}$\thanks{E-mail: vitiello@sa.infn.it} }
  \address{$^a$ Center for Theoretical Physics, Massachusetts Institute of Technology, \\
   Cambridge, MA 02139-4307, USA  \\
   and INFN, Rome, Italy}
  \address{$^b$Dipartimento di Fisica "E.R.Caianiello"
  Universit\`a di Salerno, 84081 Baronissi (SA), Italy}
  \address{$^c$INFN, Gruppo Collegato di Salerno}
  \address{$^d$INFM, Salerno, Italy}
 \date{\today}
 \maketitle
\begin{abstract}
\noindent By fully exploiting the existence of the unitarily
inequivalent representations of quantum fields, we exhibit the
entanglement between inner and outer particles, with respect to
the event horizon of a black hole. We compute the entanglement
entropy and we find that the nonunitarity of the mapping, between
the vacua in the flat and the curved frames, makes the
entanglement very robust.
\end{abstract}

\bigskip
\bigskip


\noindent Keyword(s): Entanglement and Quantum Nonlocality; QFT in
Curved Space-Times; Quantum Aspects of Black Holes, Evaporation,
Thermodynamics

\vfill

MIT-CTP 3399

\newpage

\section{Introduction}
\setcounter{equation}{0}

\noindent Black hole quantum physics nowadays is the most
promising theoretical laboratory to test fundamental ideas about
nature. Some of the most fascinating issues arising in the
struggle to understand black hole quantum physics are:  the origin
of the entropy of a black hole, the holographic paradigm, the
paradox of the information loss. All of them are related to what
seems to be a deep and fundamental problem: the counting of the
relevant degrees of freedom. Of course, all those paradoxes and
difficulties originate from the absence of a reliable quantum
theory of gravity. Hence a better understanding of them could shed
some light on the long-standing problem of quantizing gravity.

\noindent By using thermodynamical arguments, Bekenstein obtained
that the entropy of a black hole is proportional to the {\it area}
of the event horizon \cite{bekenstein},
\begin{equation}\label{bekenstein}
  S \sim {\cal A} \,,
\end{equation}
and not to the volume, as for ordinary matter. Bekenstein's
derivation left open the crucial question of the microscopic
explanation of such a result. There are, of course, many
approaches aimed to explain the microscopic origin of the black
hole entropy in several contexts. In the framework of string
theory, the microscopic derivation has been pioneered by
Strominger and Vafa \cite{vafa}, see also Ref.
\cite{Bigatti:1999dp} for a general review, and Refs.
\cite{Polchinski:1996fm,krause,schwarz,Carlip:1994gy}. On the
other hand, the problem has been faced also by considering the
value of the Euclidean action \cite{Gibbons:1976ue}, the rate of
pair creation of black holes \cite{Garfinkle:xk}, the Noether
charge associated with a bifurcating Killing horizon
\cite{Wald:nt}.

\noindent By considering a quantum scalar field in the presence of
the gravitational field, treated as a classical background
described by general relativity \cite{birrell}, Hawking showed
that quantum effects lead to thermal evaporation of black holes
\cite{HAW}, with a temperature given by
\begin{equation}\label{1a}
T = \frac{\kappa}{2\pi}\,{,}
\end{equation}
where $\kappa = (2GM)^{-1}$ is the surface gravity of the black
hole (through the paper we shall use the natural units
$\hbar=c=k_B=1$).

\noindent Hawking effect seems to be more fundamental than the
theory within which it has been derived (see for instance
\cite{Bigatti:1999dp,Visser:2001kq}), nonetheless it is surely
associated with the existence of an horizon in the space-time,
thus Schwarzschild and other background space-times have been
extensively investigated. A case in point is the Rindler
space-time \cite{rindler}, that is a flat space--time with an
horizon, associated with a uniformly accelerated observer in
Minkowski space-time \cite{Davies:1974th,Unruh:db,TAK}. A basic
point in Hawking's original proposal \cite{HAW}, is that the
evaporation of black holes through quantum radiation induces {\it
nonunitary} evolution of quantum fields. This gives raise in turn
to the intriguing puzzle that information can be lost. The
traditional quantum field theoretical approach is expected to fail
to describe these phenomena, being based on the unitarity. In
other words, if black holes' evaporation has to be described in
quantum field theory, then the request of unitarity, deeply rooted
in the conventional approach, has to be relaxed.

\noindent In this paper we shall show that the existence of the
unitarily inequivalent representations of quantum fields allows to
calculate, in a new setting, the entanglement entropy of black
holes. An unexpected consequence of the nonunitarity of the
mapping among the vacua in the flat and curved frames, is the
robustness of the entanglement between scalar quantum modes living
on the two space-like separated sides of the event horizon.

\noindent Our approach is based on the construction of an entropy
operator directly from the condensate structure of the quantum
vacuum. The main steps involved are two: first, we realize that
the vacuum state of a Minkowskian observer $|0_M \rangle$ is seen
as a condensate of entangled modes by a generic observer, and
there is no unitary transformation to disentangle $|0_M \rangle$;
second, we construct the von Neumann entropy operator from this
vacuum and investigate its properties.

\noindent Some preliminary results in this direction already
appeared while investigating more general features of field
quantization in curved background
\cite{Martellini:sm,Iorio:2001te}. Here we shall fully exploit the
entropy operator and the nonunitary nature of the entanglement to
compute the entanglement entropy of a black hole in this field
theoretical setting.

\noindent The entanglement entropy of black hole has been long
investigated
\cite{Bombelli:1986rw,Mukohyama:1996yi,Terashima:1999vw,frolov1}.
However, our present study shows that the nonunitarity of the
mapping between the vacua in the flat and curved frames is the
root of the entanglement robustness.

\noindent The paper is organized as follows: In Sec. II we recall
the main features of the quantization of a complex scalar field in
curved space-time. In Sec. III we analyze the structure of the
quantum vacuum, we show that it is a condensate of infinitely many
entangled particles living in the two space-like separated
regions, and we comment on the nonunitarity of the mapping
between the vacua in the flat and the curved frames. In Sec. IV
the entropy operator is eventually constructed and studied. We
then show in which sense its correct vacuum expectation is the
entangled entropy. The application of the obtained results to the
Schwarzschild space-time is the argument of Sec. V. Conclusions
are drawn in Sec. VI. In Appendix A we present some details of
the derivation of the entropy operator, and in Appendix B the
Rindler space-time is investigated.

\section{Quantization of Scalar Fields in Curved Space-Time}
\setcounter{equation}{0}

\noindent In this Section we recall the well known main steps of
the quantization of scalar fields in a curved background
\cite{birrell,HAW,Davies:1974th,Unruh:db,TAK,Martellini:sm,Iorio:2001te},
in order to develop the formalism to derive the entanglement and
the entropy.

\noindent Consider a complex massive scalar quantum field
$\phi(x)$ in the $D$-dimensional Minkowski space--time, with
Lagrangian density
 \begin{equation}\label{lagrangian}
   {\cal L}(\phi^*, \phi)=\partial_\mu \phi^*\partial^\mu
 \phi-m^2\phi^*\phi\,.
 \end{equation}
As usual, $\phi(x)$ can be decomposed in Minkowski modes
$\{U_k(x)\}$, orthonormal with respect to the Klein-Gordon inner
product defined as $(\phi_1, \phi_2)=i\int
\phi_1^*\stackrel{\leftrightarrow}{\partial_t}\phi_2 d^{D-1}x$.
Let us keep the $(D-1)$-volume finite, and use periodic boundary
conditions $\displaystyle{k^i_{n} = 2\pi L^{-1}_i} n_i$, with
$i=1,...,D$, $n_i$ integers, and the volume given by $L_1 \cdot
... \cdot L_D$, to write
 \begin{equation}\label{2}
 \phi(x)=\sum_k \,
 [a_{k}U_k(x)\,+\,\bar{a}_k^{\dagger}U_k^{*}(x)] \;,
 \end{equation}
with $k=(k_1, \vec{k})$. The quantum Hamiltonian is then given by
 \begin{equation} \label{hammink}
 H_M=\sum_k\,\omega_k\,
 (a_k^{\dagger}a_k\,+\,\bar{a}_k^{\dagger}\bar{a}_k) \;,
 \end{equation}
where $\omega_k= \sqrt{k_1^2+|\vec{k}|^2+m^2}$, and $a_k,
a^\dagger_k$  ($\bar{a}_k,\bar{a}_k^{\dagger}$) are the
annihilation and creation operators, respectively, for particles
(antiparticles). They act on the Hilbert-Fock space $\cal H$, the
Minkowski vacuum being defined by
 \begin{equation}\label{4}
 a_k|0_M\rangle\,=\,\bar{a}_k|0_M\rangle\,=0, \quad \forall \,k\,{.}
 \end{equation}
and satisfy the usual canonical commutation relations (CCRs)
 \[
 [a_k, a_{k'}^\dag]=\delta_{kk'}\,, \quad [\bar{a}_k,
 \bar{a}_{k'}^\dag]=\delta_{kk'}\,.
 \]
Let us now recall the key features of the quantization of the same
scalar field $\phi(x)$ in curved space-times. One expands the
field in terms of the complete set (with respect to Klein-Gordon
product) of functions $\{u_p^{(\sigma)}(x)\}$, where $p=(\Omega,
{\vec k})$, and the symbol $\sigma = \pm$ takes into account the
fact that the space-time has an horizon, so that it is divided
into two causally disconnected regions: the internal and external
regions of the Schwarzschild geometry, the two Rindler wedges,
etc.. The functions $\{u_p^{(\sigma)}(x)\}$ are derived by solving
the Klein-Gordon equation in the coordinates of space-time under
consideration. Thus, one writes
\begin{equation}\label{phiR}
  \phi(x)= \sum_\sigma \sum_p \, \left[
  b_p^{(\sigma)}u_p^{(\sigma)}(x)+
  {\bar b}_p^{(\sigma)\, \dagger}u_p^{(\sigma)\, *}(x)
  \right]\,,
\end{equation}
where the operators $b_p^{(\sigma)}$ and ${\bar b}_p^{(\sigma)}$
are assumed to satisfy the usual CCRs.

\noindent By introducing the operators \cite{TAK}
\begin{equation}\label{d-op}
  d^{(\sigma)}_p=\sum_{k_1} {\cal
  P}_\Omega^{(\sigma)}(k_1)\, a_{k_1{\vec k}}\,,
\end{equation}
and similarly for ${\bar d}^{(\sigma)}_p$ in terms of ${\bar
a}_k$, where $\{{\cal P}_\Omega^{(\sigma)}(k_1)\}$ is a complete
set of orthogonal functions, the operators $b_p^{(\sigma)}$, and
${\bar b}_p^{(\sigma)}$ can be expressed in terms of the Bogolubov
transformations \cite{birrell,TAK}
 \begin{eqnarray}
 b_p^{(\sigma)} & = & d_p^{(\sigma)} \cosh \epsilon (p)
       + {\bar d}_{\tilde p}^{(-\sigma) \dagger} \sinh \epsilon
       (p) = G(\epsilon)d_p^{(\sigma)}G^{-1}
 (\epsilon)\,, \label{b1} \\
 {\bar b}_{\tilde p}^{(-\sigma) \dagger}  & = &
 d_p^{(\sigma)} \sinh \epsilon (p)
 + {\bar d}_{\tilde p}^{(-\sigma) \dagger} \cosh \epsilon
 (p) = G(\epsilon)\bar{d}_{\tilde p}^{(-\sigma)}G^{-1}
 (\epsilon) \,, \label{b2}
 \end{eqnarray}
where $\tilde{p} = (\Omega, -\vec{k})$, and the generator of the
transformations is
 \begin{equation}\label{eq16}
 G(\epsilon) = \exp \left\{  \sum_{\sigma} \sum_p \epsilon(p)
 \left[ d_p^{(\sigma)}\bar{d}_{\tilde p}^{(-\sigma)} -
 d_p^{(\sigma)\,\dagger}\bar{d}_{\tilde
 p}^{(-\sigma)\,\dagger} \right] \right\} \,{.}
 \end{equation}
At finite volume $G(\epsilon)$ is a unitary operator:
 \[
 G^{-1}(\epsilon)=G(-\epsilon)=G^{\dagger}(\epsilon)\,.
 \]
The canonical operators $d^{(\sigma)}_p$ and ${\bar
d}^{(\sigma)}_p$ annihilate the Minkowski vacuum
$|0^{(+)}_M\rangle \otimes |0^{(-)}_M\rangle$. On the other hand,
the operators $b_p^{(\sigma)}$, and ${\bar b}_p^{(\sigma)}$
annihilate the vacuum
\begin{equation}\label{vacb}
  \vert 0^{(+)} (\epsilon) \rangle \otimes  \vert 0^{(-)} (\epsilon) \rangle
  = G (\epsilon) \left[\vert 0_M^{(+)} \rangle \otimes  \vert 0_M^{(-)} \rangle \right] \,.
\end{equation}
In what follows we shall keep the short-hand notation: $\vert 0_M
\rangle \equiv \vert 0_M^{(+)} \rangle \otimes \vert 0_M^{(-)}
\rangle$, and $\vert 0 (\epsilon) \rangle \equiv \vert 0^{(+)}
(\epsilon) \rangle \otimes \vert 0^{(-)} (\epsilon) \rangle$.

\noindent The parameter $\epsilon$ in Eqs. (\ref{b1}) and
(\ref{b2}) is given by
\begin{equation}\label{boson-stat}
 \sinh \epsilon (p)= \frac{1}{(e^{\Omega/T}-1)^{1/2}}\,,
\end{equation}
where $T$ is related to the surface gravity of black holes, in the
case of Schwarzschild geometry \cite{HAW}, or to the acceleration,
in the case of Rindler geometry \cite{Unruh:db}. One can also show
that the total Hamiltonian is given by
\cite{birrell,HAW,Davies:1974th,Unruh:db,TAK,Iorio:2001te,umezawa}
\begin{eqnarray}
  H_{\epsilon} & = & \sum_{\sigma}\sum_p \sigma \Omega \,[ b_p^{(\sigma)
 \dagger} b_p^{(\sigma)} +
 \bar{b}_{\tilde{p}}^{(\sigma)}
 \bar{b}_{\tilde{p}}^{(\sigma)\dagger}] \nonumber \\
 &=& H^{(+)}(\epsilon) - H^{(-)}(\epsilon) \,{,}
 \label{hamrin}
 \end{eqnarray}
where we chose not to normal-order, as some considerations on the
entropy will be clearer in this setting (see Sec. V). Let us also
note that the vacuum $|0 (\epsilon) \rangle$ does depend on $T$,
hence depends on the physical parameters that characterize the
background.

\noindent The Bogolubov transformations, Eqs. (\ref{b1}) and
(\ref{b2}), relate the vectors of ${\cal H} = {\cal H}^{(+)}
\otimes {\cal H}^{(-)}$ to the vectors of another Hilbert--Fock
space ${\cal H}_{\epsilon} = {\cal H}^{(+)}_{\epsilon} \otimes
{\cal H}^{(-)}_{\epsilon}$ labelled by $\epsilon$. In our approach
the structure of the vacuum, and of course of the Hilbert space,
is crucial. We shall dedicate the next Section to the
investigation of their physical properties.

\section{Vacuum Structure and Entanglement}
\setcounter{equation}{0}

\noindent The relation between the spaces ${\cal H}$ and ${\cal
H}_{\epsilon}$ is established by the generator $G(\epsilon)$:
${\cal H} \to {\cal H}_{\epsilon}$, or by its inverse
$G^{-1}(\epsilon)$: ${\cal H}_\epsilon \to {\cal H}$. The physical
meaning of this freedom is that we can arbitrarily choose to
express Minkowskian quantities in terms of generic
$\epsilon$-quantities, or the other way around. We choose, for
instance, to express the Minkowskian vacuum in terms of the
generic $\epsilon$-vacuum
 \begin{equation}\label{19}
 |0_M\rangle \, = \,G^{-1}(\epsilon)|0 (\epsilon) \rangle \,.
 \end{equation}
This relation holds only at finite degrees of freedom, i.e. finite
volume. Note that $G(\epsilon)$ is an element of\footnote{To be
more precise, for each mode, we have the direct product of {\it
two} two-boson realizations of $SU(1,1)$: $[T^i_+,T^i_-]=
-2T^i_0$, $[T^i_0,T^i_\pm]=\pm T^i_\pm$, $i=1,2$, where
$[T^1,T^2]=0$, for all the generators $T$. This is seen by
defining $T^1_+ = d^{(+) \dagger} \bar{d}^{(-) \dagger}$, $T^1_-=
d^{(+)} \bar{d}^{(-)}$, $T^1_0 = \frac{1}{2} (d^{(+) \dagger}
d^{(+)} + \bar{d}^{(-) \dagger} \bar{d}^{(-)})$, while $T^2_+$,
$T^2_-$, $T^2_0$ are obtained by replacing $(+) \leftrightarrow
(-)$.} $SU(1, 1) \times SU(1, 1)$. Of course the same structure
arises by writing $G^{-1}(\epsilon)$ in terms of the $b$s, all one
has to do is to replace $d \to b$ in the Eq. (\ref{eq16}). Thus by
using the Gaussian decomposition, the Minkowski vacuum can be
formally expressed as a $SU(1,1) \times SU(1,1)$ generalized
coherent state \cite{PER} of Cooper-like pairs
 \begin{equation}\label{26}
 |0_M\rangle =\frac{1}{Z}\,\exp\left[{\sum_{\sigma} \sum_p \;\tanh\epsilon (p)
 b_p^{(\sigma)\dagger} \bar{b}_{\tilde
 p}^{(-\sigma)\dagger}}\right]\, |0(\epsilon)\rangle\,{,}
 \end{equation}
where $Z= \prod_p\;\cosh^2\epsilon(p)$.

\noindent In the continuum limit in the space of momenta, i.e. in
the infinite-volume limit, the number of degrees of freedom
becomes uncountable infinite, hence we have
\begin{eqnarray}
 \langle 0(\epsilon)| 0_M\rangle \to 0 & &\quad {\rm as}
 \quad  V\to\infty, \quad \forall
 \epsilon
 \label{eq27} \\
  \langle 0(\epsilon)|0(\epsilon^{\prime})\rangle\to 0 & &\quad {\rm as} \quad
   V\to\infty, \quad \forall
  \epsilon, \epsilon^{\prime}, \epsilon\ne \epsilon^{\prime}\,{,}
  \label{eq28}
\end{eqnarray}
where $V$ is the volume of the whole $(D-1)$-dimensional space.
This means that the Hilbert spaces ${\cal H}$ and ${\cal
H}_\epsilon$ become unitarily inequivalent in the continuum limit.
In this limit $\epsilon$ labels the set $\{H_\epsilon, \forall
\epsilon\}$ of the infinitely many unitarily inequivalent
representations of the CCRs \cite{umezawa,strocchi}.

\noindent Let us now discuss the entanglement of the vacuum $|0_M
\rangle$ in (\ref{26}), that we rewrite in the following
convenient form
\begin{equation}
  |0_M \rangle = \frac{1}{Z} \left[ |0(\epsilon)\rangle  + \sum_p \;\tanh\epsilon (p)
  \left( | 1^{(+)}_p , \bar{0} \rangle \otimes |0, \bar{1}^{(-)}_{\tilde p} \rangle
  + | 0, \bar{1}^{(+)}_{\tilde p} \rangle \otimes |1^{(-)}_{ p} , \bar{0} \rangle
  \right)  + \dots \right] \,, \label{expans-vacuum}
\end{equation}
where, we denote by $|n^{(\sigma)}_p, \bar{m}^{(\sigma)}_{\tilde p
}\rangle$ a state of $n$ particles and $m$ antiparticles in
whichever sector $(\sigma)$. Note that for the generic $n^{\rm
th}$ term, the state $| n^{(\sigma)}_p , \bar{0} \rangle \equiv
|1^{(\sigma)}_{p_1}, \ldots, 1^{(\sigma)}_{p_n}, \bar{0} \rangle$,
and similarly for antiparticles.

\noindent By introducing a well-known notation, $\uparrow$ for a
particle, and $\downarrow$ for an antiparticle, the two-particle
state in (\ref{expans-vacuum}) can be written as
\begin{equation}\label{enta-onestate}
| \uparrow^{(+)} \rangle \otimes | \downarrow^{(-)} \rangle + |
\downarrow^{(+)} \rangle \otimes | \uparrow^{(-)} \rangle \,,
\end{equation}
which is an entangled state of particle and antiparticle living in
the two causally disconnected regions $(\pm)$. The generic $n^{\rm
th}$ term in (\ref{expans-vacuum}) shares exactly the same
property as the two-particle state, but this time the $\uparrow$
describes a {\it set} of $n$ particles, and $\downarrow$ a {\it
set} of $n$ anti-particles. The mechanism of the entanglement,
dynamically induced by gravitational effects, takes place at all
orders in the expansion, always by grouping particles and
antiparticles into two sets. Thus the whole vacuum $|0_M\rangle$
is an infinite superposition of entangled states\footnote{A
similar structure also arises in the temperature-dependent vacuum
of Thermo-Field Dynamics \cite{umezawa}, see also Ref.
\cite{cinesi}.}
\begin{equation}\label{enta-series}
  |0_M\rangle = \sum_{n=0}^{+\infty} \sqrt{W_n} |{\rm Entangled} \rangle_n
  \,,
\end{equation}
where
\begin{equation}\label{wn}
  W_n = \prod_p
  \frac{\sinh^{2n_p}\epsilon(p)}{\cosh^{2(n_p+2)}\epsilon(p)}\,,
\end{equation}
with
\begin{equation}\label{wn1}
0< W_n <1 \quad {\rm and} \quad  \sum_{n=0}^{+\infty} W_n = 1 \,.
\end{equation}
Details of the computations can be found in the Appendix A.

\noindent Of course, the probability of having entanglement of two
sets of $n$ particles and $n$ antiparticles is $W_n$. At finite
volume, being $W_n$ a decreasing monotonic function of $n$, the
entanglement is suppressed for large $n$. It appears then, that
only a finite number of entangled terms in the expansion
(\ref{enta-series}) is relevant. Nonetheless this is only true at
finite volume (the quantum mechanical limit), while the
interesting case occurs in the infinite volume limit, which one
has to perform in a quantum field theoretical setting.

\noindent The entanglement is generated by $G(\epsilon)$, where
the scalar field modes in one sector $(\sigma)$ are coupled to the
modes in the other sector $(-\sigma)$ via the parameter
$\epsilon(p)$, which takes into account the effects of the
background gravitational field (environment) \cite{Iorio:2001te}.
We stress that the origin of the entanglement {\it is} the
environment, in contrast with the usual quantum mechanical view,
which attributes to the environment the loss of the entanglement.
In the present treatment such an origin for the entanglement makes
it quite robust.

\noindent One further reason for the robustness is that this
entanglement is realized in the limit to the infinite volume once
and for all, since then there is no unitary evolution to
disentangle the vacuum: at infinite volume one cannot "unknot the
knots". Such a nonunitarity is only realized when {\it all} the
terms in the series (\ref{enta-series}) are summed up, which
indeed happens in the $V\to \infty$ limit.

\noindent Summarizing, the interaction with the environment and
nonunitarity, are the basis for the generation and the stability
of the entanglement. These two features are entirely due to the
nature of the quantum field theoretical vacuum.

\section{The Entropy Operator}
\setcounter{equation}{0}

\noindent It is now matter of constructing thermodynamical
quantities out of the condensate structure of the entangled vacuum
$|0_M \rangle$. By using Eqs. (\ref{b1}) and (\ref{b2}) one
obtains that the number of modes of the type $b_p^{(\sigma)}$ in
$|0_M\rangle$ is given by
 \begin{equation}\label{25}
 {\cal N}^{(\sigma)}_{b} \equiv \langle 0_M|b_p^{(\sigma)
 \dagger} b_p^{(\sigma)}|0_M\rangle =
 \sinh^2\epsilon(p) \;, \quad \sigma = \pm \;,
 \end{equation}
and similarly for the modes of type $\bar{b}_{\tilde
p}^{(\sigma)}$.

\noindent Algebraic manipulations allow to recast $|0_M \rangle$
in Eq. (\ref{26}) in the form
\begin{eqnarray}\label{vacua+entropy}
  |0_M\rangle &=& e^{-S^{(+)}(\epsilon)/2}\,
        e^{\sum_{p, \sigma}b^{(\sigma)\, \dagger}_p\,
        {\bar b}^{(-\sigma)\, \dagger}_{\tilde
        p}}|0(\epsilon)\rangle
        \\
        &=& e^{-S^{(-)}(\epsilon)/2}\,
        e^{\sum_{p, \sigma}b^{(\sigma)\, \dagger}_p\,
        {\bar b}^{(-\sigma)\, \dagger}_{\tilde
        p}}|0(\epsilon)\rangle \,,
        \nonumber
\end{eqnarray}
where $S^{(+)}(\epsilon)$ and $S^{(-)}(\epsilon)$ are given by
 \begin{eqnarray}\label{S+}
 S^{(+)}(\epsilon)&=& {\cal S}^{(+)}(\epsilon) + \bar{\cal
 S}^{(+)}(\epsilon) \\
 &=& - \sum_p [b_p^{(+) \dagger} b_p^{(+)}
  \ln\sinh^2\epsilon (p)
 - b_p^{(+)} b_p^{(+)\dagger} \ln\cosh^2\epsilon
 (p) \nonumber \\
 & & + (b \to \bar{b})] \,, \nonumber
 \end{eqnarray}
 \begin{eqnarray}\label{S-}
 S^{(-)}(\epsilon)&=& {\cal S}^{(-)}(\epsilon) + \bar{\cal
 S}^{(-)}(\epsilon) \\
 &=& - \sum_p [b_p^{(-) \dagger} b_p^{(-)} \ln\sinh^2\epsilon (p)
 - b_p^{(-)} b_p^{(-)\dagger} \ln\cosh^2\epsilon
 (p) \nonumber \\
 & & + (b \to \bar{b})] \,. \nonumber
 \end{eqnarray}
The formulae (\ref{vacua+entropy}) will be proved in some details
in the Appendix A.

\noindent One can easily convince himself that
$S^{(\sigma)}(\epsilon)$, $\sigma = \pm$, is nothing else than the
von Neumann entropy
 \begin{equation}\label{vonNeuE}
 S = - {\cal N} \ln {\cal N} \,,
 \end{equation}
where ${\cal N}$ is the number of microscopic states.  At the
origin of this entropy there are the vacuum fluctuations of the
quantum states, which have a thermal character for different
observers related to the Minkowski observer through a
diffeomorphism. By counting the number of occupation states in the
vacuum $|0_M\rangle$ with the number operator for particles
$N^{(\sigma)}_{b} \equiv b_p^{(\sigma) \,\dagger} b_p^{(\sigma)}$,
we must subtract those occupation states counted by the operator
$\bar{b}_{\tilde p}^{(\sigma)} \bar{b}_{\tilde p}^{(\sigma)\,
\dagger} = 1 + N^{(\sigma)}_{\bar{b}}$, where
$N^{(\sigma)}_{\bar{b}}$ is the number operator for the
antiparticles. This accounts for our definitions (\ref{S+}) and
(\ref{S-}) of the entropy operators
 \begin{equation} \label{23}
   S^{(\sigma)}(\epsilon) =  - \sum_p \left[N^{(\sigma)}_{b}
 \ln {\cal N}^{(\sigma)}_{b} - \left(1 + N^{(\sigma)}_{\bar b}\right) \ln \left(1 +
 {\cal N}^{(\sigma)}_{\bar{b}}\right)+ (b \to \bar{b})\right] \,.
 \end{equation}
We note that in terms of the coefficients $W_n$ in Eq. (\ref{wn})
we have
\[
\langle0_M\vert S^{(\sigma)}(\epsilon) \vert0_M\rangle = - \sum_{n
\geq 0} W_n \ln W_n \,.
\]
We emphasize that the operator $S^{(+)}(\epsilon) = {\cal
S}^{(+)}(\epsilon) + \bar{\cal S}^{(+)}(\epsilon)$ is the sum of
the entropy operators for the boson gas of particles and
antiparticles in the sector $(+)$, similarly for
$S^{(-)}(\epsilon)$ in the sector $(-)$. The total entropy
operator is given by
\begin{equation}\label{Stotale}
S_\epsilon = S^{(+)}(\epsilon) - S^{(-)}(\epsilon) \,,
\end{equation}
and, as the Hamiltonian (\ref{hamrin}), it is the {\it difference}
of the two operators. The Bogolubov transformations leave
$S_\epsilon$ invariant, $[S_\epsilon, G(\epsilon)]=0$, hence
$S_\epsilon |0_M\rangle = 0$. This means that one can arbitrarily
choose one of the two sectors, $\sigma = \pm $, to ``measure" the
correspondent entropy $S^{(\pm)}(\epsilon)$ relative to the ground
state $|0_M\rangle$. Let us work in the sector $\sigma = +$.

\noindent Now we want to show in which sense this entropy is an
entanglement entropy. When one computes the vacuum expectation
value $\langle 0(\epsilon)| S^{(+)}(\epsilon)|0(\epsilon)\rangle$
the result is
 \[
 \langle
0(\epsilon)| S^{(+)}(\epsilon)|0(\epsilon)\rangle =
-2\sum_{\Omega, \vec k} \ln \cosh^2\epsilon(\Omega) \,,
 \]
which is divergent due to the infinite sum on the momenta
$\sum_{\vec k}$. Nevertheless, if the entropy operator is
normal-ordered, the term $\ln \cosh^2\epsilon(\Omega)$ is removed
and the expectation value of the entropy {\it vanishes}.

\noindent Nonetheless with $S^{(+)}(\epsilon)$ we have to look at
the entropy of the Minkowski vacuum $|0_M\rangle$, and not of
$|0(\epsilon)\rangle$. The generic observer sees the Minkowski
vacuum as an entangled condensate. Such an observer then will
measure $\langle 0_M| S^{(+)}(\epsilon)|0_M\rangle$. The result of
such a "cross measurement" gives
\begin{equation}
\label{entropy-cross} \langle 0_M|S^{(+)}(\epsilon)|0_M\rangle = -
2 \sum_{\Omega, \vec k} \left[\sinh^2 \epsilon(\Omega) \ln \sinh^2
\epsilon(\Omega) - \cosh^2 \epsilon(\Omega) \ln \cosh^2
\epsilon(\Omega)\right] \,,
\end{equation}
where Eq. (\ref{25}) has been used. We notice that again there is
a divergence due to $\sum_{\vec k}$, but this time even if the
expression is normal-ordered the result is {\it never zero} (cfr.
e.g. Sec V).

\noindent The physical meaning of such occurrence is that the
entanglement is dynamically generated via the interaction with the
gravitational background, as already observed. The entanglement is
only seen if the generic observer measures on the Minkowski
vacuum. It is precisely in this sense that we say that the black
hole entropy is the entanglement entropy.

\noindent As already observed, in the $V \to \infty$ limit the
generator $G(\epsilon)$ does not exist, but the whole structure
described above survives. What one loses in that limit is the
possibility to generate the mapping between the states of the two
frames, the Minkowski and the generic one. In some sense, the
''symmetry'' in writing $\langle 0_M| S^{(+)}(\epsilon)|0_M\rangle
= \langle 0(\epsilon)|G(\epsilon) S^{(+)}(\epsilon)
G^{-1}(\epsilon) |0(\epsilon)\rangle =\langle 0(\epsilon)|
S^{(+)}_M|0(\epsilon)\rangle $ is lost.

\noindent For the sake of completeness we conclude this Section by
noticing that the "free-energy" can also be introduced
\cite{umezawa}
 \begin{equation}\label{22}
  {\cal F}^{(+)}(\epsilon)\equiv \langle 0_M|H^{(+)} (\epsilon)
  -\frac{1}{\beta} \, S^{(+)}(\epsilon)|0_M\rangle \,{.}
 \end{equation}
It is interesting to note that by looking for the values of
$\epsilon (p)$ making ${\cal F}^{(+)}(\epsilon)$ stationary, and
considering negligible variations of $\beta$ with respect to
$\epsilon$ one obtains
 \begin{equation}\label{24}
 \beta\Omega = - \ln\tanh^2\epsilon (p) \Leftrightarrow
 \sinh^2\epsilon (p)= \frac{1}{e^{\beta\Omega}-1}\,{.}
 \end{equation}
From Eq. (\ref{25}), it follows that
 \begin{equation}\label{Ndist}
 {\cal N}^{(+)}_{b} = \frac{1}{e^{\beta\Omega}-1}\,,
 \end{equation}
and similarly for ${\cal N}^{(+)}_{\bar{b}}$. In this way one
consistently recovers (\ref{boson-stat}), which is the Bose
distribution provided that $T=\beta^{-1}$ is identified with the
temperature.

\section{Applications to the Schwarzschild Space--Time}
\setcounter{equation}{0}

\noindent We derive now the thermodynamical properties of black holes
described by the Schwarzschild geometry
 \begin{equation}
 ds^2=\left( 1-\frac{2GM}{r} \right) dt^2-
 \left( 1-\frac{2GM}{r} \right)^{-1}dr^2 -
 r^2(d\theta^2+\sin^2\theta d\varphi^2)\,,
 \label{SCHW}
 \end{equation}
where the space-time is taken to be 4-dimensional. The event
horizon is given by $r_S = 2GM$, and the Bekenstein-Hawking
temperature (\ref{1a}) is $T \simeq r^{-1}_S = (2GM)^{-1}$.

\noindent We want to calculate the finite part of the entropy in
Eq. (\ref{entropy-cross}) by moving to the continuum limit, taken
to be formally identical to the Minkowskian one
 \begin{equation}\label{continuum}
\sum_{\Omega, \vec k} \to \frac{V}{(2\pi)^3}\int_0^\infty
d\Omega\int d^2 k\,,
 \end{equation}
where $V$ is the 3-dimensional volume of the whole space-time.

\noindent The entropy {\it density} is
\begin{eqnarray}\label{dens-entropy}
\langle s^{(+)}(\epsilon)\rangle_M &\equiv&
 \frac{\langle 0_M|S^{(+)}(\epsilon)|0_M\rangle}{V} \\
 &=& \frac{-2}{(2\pi)^3}\int_0^{\infty}d \Omega \left[\sinh^2 \epsilon(\Omega)
\ln \sinh^2\epsilon(\Omega) - \cosh^2 \epsilon(\Omega)\ln
 \cosh^2\epsilon(\Omega)\right]\int d^2 k \,{.} \nonumber
 \end{eqnarray}
By expressing $\epsilon(\Omega)$ as a function of $\Omega$ through
Eq. (\ref{boson-stat}), one can compute the integral in  Eq.
(\ref{dens-entropy}) to obtain (see Eqs. 3.411-1 and 4.223-2 of
Ref. \cite{ryzhik})
\begin{equation}\label{dens-entr-int}
 \langle s^{(+)}(\epsilon)\rangle_M
 = \frac{\pi^2 T}{3(2\pi)^3}  \int d^2 k \,{,}
\end{equation}
which is divergent, as expected. We cannot remove this infinity by
normal ordering the entropy operator. Following the recipe adopted in 
quantum gravity we use a cutoff $k_C$ of the order of the Planck 
momentum $k_C \simeq k_P=l_P^{-1}=G^{-1/2}$. 
Our entropy density is then given by
\begin{equation}\label{dens-entr-cutoff}
 \langle s^{(+)}(\epsilon)\rangle_M
 = \frac{k^2_C T}{24 \pi}  \,{.}
\end{equation}
Being the proper volume in the Schwarzschild geometry
\[
V_{\rm prop} = \int \sqrt{-g_{rr}
g_{\theta\theta}g_{\varphi\varphi}}\, dr d\theta d\varphi
\]
only defined for $r > r_S$, we now have to compute the entropy for
the spherical shell of radii $r_S$ and $r_S + \delta$. The volume
of the shell is given by
\begin{equation}\label{vshell}
  {\cal V} = 4 \pi r_S^3 \int_1^{1 + h}\frac{x^{5/2}}{\sqrt{x-1}}
  \; dx \propto r_S^3  \,,
\end{equation}
where $h = \delta / r_S $ is chosen by requiring the numerical
factor of proportionality to be ${\cal O}(1)$. Since $k_C \lesssim
k_P$, and recalling that the Bekenstein-Hawking temperature is
$T\sim r_S^{-1}$, we obtain the upper bound for the entropy
\begin{equation}\label{entropy1}
 \langle S^{(+)}(\epsilon)\rangle_M = {\cal V} \; \langle s^{(+)}(\epsilon)\rangle_M
 \lesssim \frac{\cal A}{l_P^2} \,{.}
\end{equation}
Hence the entropy is proportional to the horizon area ${\cal A}$
of the black hole and is bounded from above. In the Appendix B
this analysis will be applied to the Rindler space-time.

\noindent It is an interesting question to investigate the relation
between the derivation of (\ref{entropy1}) and the holographic
principle.

\section{Conclusions}

\noindent We proposed a new method to obtain the entanglement
entropy of black holes, based on the relevant nonunitary features of 
quantum fields. This was done for the Schwarzschild and Rindler (Appendix
B) space-times.

\noindent Hawking's original proposal \cite{HAW} that the
evaporation of black holes through quantum radiation induces
nonunitary evolution of quantum fields was reanalysed. Despite the
fact that a description in terms of
the traditional (fully unitary) quantum field theoretical approach 
is then expected to fail, we have shown that
the unitarily inequivalent representation of quantum
fields allow to calculate, in a new setting, the entanglement
entropy of black holes. 

\noindent We have also shown that the entanglement
between inner and outer particles, with respect to the event
horizon, is very robust due to the nonunitary nature of
the mapping between the vacua in the flat and the curved frames.

\noindent Our results are obtained assuming a black hole at
thermodynamical equilibrium. The approach presented in this paper
can be applied to a wide range of problems related to black holes'
quantum physics. We shall further exploit this formalism in future
works.

\acknowledgments

\noindent A.I. thanks Roman Jackiw for interesting comments. A.I. and G.V. 
acknowledge COSLAB Network European Science Foundation. This work is 
supported in part by funds provided by the U.S. Department of Energy 
(D.O.E.) under cooperative research agreement DF-FC02-94ER40818.

\appendixa

\noindent In this Appendix we want to derive the expressions
(\ref{vacua+entropy}) and (\ref{enta-series}) for the vacuum
$|0_M\rangle$. We first derive a couple of useful relations.
Recalling that $S^{(\sigma)}(\epsilon) = {\cal
S}^{(\sigma)}(\epsilon) + \bar{\cal S}^{(\sigma)}(\epsilon)$, we
have
\begin{eqnarray}
  e^{-S^{(\sigma)}(\epsilon)/2}\, b^{(\sigma)\dagger}_p
 e^{S^{(\sigma)}(\epsilon)/2} &=& b^{(\sigma)\dagger}_p + \frac{1}{2}
 [b^{(\sigma)\dagger}_p,{\cal S}^{(\sigma)}(\epsilon)] + \frac{1}{8}
 [[b^{(\sigma)\dagger}_p,{\cal S}^{(\sigma)}(\epsilon)],{\cal S}^{(\sigma)}(\epsilon)] + \ldots
 \nonumber \\
 &=& b^{(\sigma)\dagger}_p \left( 1+ \ln \tanh \epsilon(p) +
 \frac{1}{2} (\ln \tanh \epsilon(p))^2 + \ldots \right)
\nonumber \\
&=& b^{(\sigma)\dagger}_p \exp \{ \ln \tanh \epsilon(p) \} =
b^{(\sigma)\dagger}_p \tanh \epsilon(p) \,, \label{a1}
\end{eqnarray}
where $\sigma = \pm$. Similarly one can prove the same relation
for the antiparticles by using the $\bar{\cal
S}^{(\sigma)}(\epsilon)$ in $S^{(\sigma)}(\epsilon)$:
 \begin{equation}
 e^{-S^{(\sigma)}(\epsilon)/2}\, \bar{b}^{(\sigma)\dagger}_p
 e^{S^{(\sigma)}(\epsilon)/2}
 =\tanh \epsilon(p) \,\bar{b}^{(\sigma)\dagger}_p \,.
\end{equation}
Furthermore, one sees that
\begin{eqnarray}\label{S0+}
e^{-S^{(\sigma)}(\epsilon)/2}\,|0(\epsilon)\rangle & =&
             \exp \left\{\frac{1}{2} \sum_p \left[\left(N_b^{(\sigma)}\ln
             \tanh^2\epsilon(p)-\ln\cosh^2\epsilon(p)\right)+(b\to {\bar
             b})\right]\right\}\,|0(\epsilon)\rangle \\
             &=&\exp\{\sum_p\ln\tanh\epsilon(p)\, N_b^{(\sigma)} \}
             \exp\{-\sum_p\ln\cosh\epsilon(p)\}\times (b\to {\bar b})\,|0(\epsilon)\rangle
             \,.  \nonumber
\end{eqnarray}
By noting that
\begin{eqnarray}\label{S1+}
 \exp\{\sum_p\ln\tanh\epsilon(p)\, N_b^{(\sigma)} \}\,|0(\epsilon)\rangle &=& |0(\epsilon)\rangle\,, \\
 \exp\{-\sum_p\ln\cosh\epsilon(p)\}\,|0(\epsilon)\rangle &=&\prod_p
 \cosh^{-1}\epsilon(p)\,|0(\epsilon)\rangle\,, \nonumber
  \end{eqnarray}
and similarly for antiparticles, one gets
\begin{equation}\label{relaz2-A}
 e^{-S^{(\sigma)}(\epsilon)/2}\,|0(\epsilon)\rangle =
 \left[\prod_p\,\cosh^{-2} \epsilon(p)\right]|0(\epsilon)\rangle\,.
\end{equation}
Now we are in the position to prove Eq. (\ref{vacua+entropy}). Eq.
(\ref{26}) can be rewritten as
\begin{equation}\label{vac1A}
  |0_M\rangle = \prod_\sigma\prod_p \left( 1+\ldots +
  \frac{1}{n!}\,\tanh^n\epsilon(p)\,
  b_p^{(\sigma)\dagger} \bar{b}_{\tilde
 p}^{(-\sigma)\dagger}+\ldots \right)\,
 \left[\prod_p\,\cosh^{-2} \epsilon(p)\right] |0(\epsilon)\rangle\,{,}
\end{equation}
and by using Eqs. (\ref{relaz2-A}) and (\ref{a1}), one obtains
\begin{equation}\label{vacfin}
 |0_M\rangle = \left(e^{-S^{(+)}/2}e^{\sum_\sigma\sum_p b_p^{(\sigma)\dagger} \bar{b}_{\tilde
 p}^{(-\sigma)\dagger}}e^{S^{(+)}/2}\right)\left[\,e^{-S^{(+)}/2}|0(\epsilon)\rangle \right]
 \,,
\end{equation}
similarly with $S^{(-)}$, q.e.d.\,.

\noindent Let us now prove that the coefficients $W_n$ in Eq.
(\ref{enta-series}) are given by Eq. (\ref{wn}). At this end, we
recast the vacuum (\ref{vacua+entropy}) in the form
\begin{eqnarray}\label{vac2A}
  |0_M\rangle &=&e^{-S^{(+)}/2}\sum_{n_p=0}^\infty \left[
  |n_p^{(+)}, {\bar 0}\rangle \otimes |0, {\bar n}_{\tilde
  p}^{(-)}\rangle + |0, {\bar n}_{\tilde
  p}^{(+)}\rangle \otimes |n_p^{(-)}, {\bar 0}\rangle\right] \\
   &\equiv & e^{-S^{(+)}/2} \sum_{n_p=0}^\infty \sum_{\sigma=\pm} |n_p^{(\sigma)};{\bar n}_{\tilde
   p}^{(-\sigma)}\rangle \nonumber \,,
\end{eqnarray}
where, according to the notation introduced after Eq.
(\ref{expans-vacuum}) and in Eq. (\ref{enta-series}), $n={\bar
n}$, and
 \[
 \sum_{\sigma=\pm} |n_p^{(\sigma)}; {\bar n}_{\tilde
   p}^{(-\sigma)}\rangle \equiv |{\rm Entangled}\rangle_{n_p}\,.
 \]
Hence,
\begin{eqnarray}\label{vac3A}
 |0_M\rangle &=& \sum_{n_p=0}^\infty e^{\sum_{p'}[n_{p'}\ln \sinh\epsilon(p')-
 (1+n_{p'})\ln\cosh\epsilon(p')]}
 \sum_{\sigma=\pm} |n_p^{(\sigma)}; {\bar n}_{\tilde
   p}^{(-\sigma)}\rangle  \\
  &=& \sum_{n_p=0}^\infty\prod_p
  \tanh^{n_p}\epsilon(p)\cosh^{-1}\epsilon(p)\sum_{\sigma=\pm} |n_p^{(\sigma)}; {\bar n}_{\tilde
   p}^{(-\sigma)}\rangle \nonumber \\
   &=& \sum_{n_p=0}^\infty \sqrt{W_{n_p}}\sum_{\sigma=\pm} |n_p^{(\sigma)}; {\bar n}_{\tilde
   p}^{(-\sigma)}\rangle\,, \nonumber
\end{eqnarray}
which proves Eq. (\ref{wn}).

\appendixb

\noindent In this Appendix we apply the formalism developed in
Secs. IV and V to the Rindler space-time \cite{rindler},
corresponding to an observer moving with constant acceleration
$a$. This space-time is described by the line element (see for
example \cite{birrell})
\begin{equation}\label{rindler}
  ds^2=e^{2a\xi}(d\tau^2-d\xi^2)-dy^2-dz^2\,,
\end{equation}
which reduces to Minkowski space-time letting
 \begin{equation}
 t=\frac{e^{a\xi}}{a}\sinh a\tau\,, \quad x=\frac{e^{a\xi}}{a}\cosh
 a\tau\,.
 \end{equation}
This metric covers a portion of Minkowski space-time with $x>\vert
t\vert$. The boundary planes are determined by $x\pm t=0$
\cite{rindler}. Davies \cite{Davies:1974th} and Unruh
\cite{Unruh:db} have shown that the vacuum state for an inertial
observer is a canonical ensemble for the Rindler observer. The
temperature $T_R$ characterizing  this ensemble is related to the
acceleration of the observer by the relation
\begin{equation}\label{1}
T_R\,=\,\frac{a}{2\pi} \;.
\end{equation}
This is the thermalization theorem, in a nutshell (for a review
see \cite{TAK}).

\noindent The proper volume is given by
 \begin{equation}
 V_{\rm prop}= \int \sqrt{g_{\xi\xi} g_{yy} g_{zz}} \, d\xi dy dz =
 \int_{-\infty}^0 e^{a\xi}\,d\xi \int dy dz = \frac{\cal A}{a} \,,
 \end{equation}
where ${\cal A} = \int dy dz$ is the area of a surface of constant
$\xi$ and $\tau$. The entropy density is computed for a cutoff on
the momenta $k_C \lesssim l_P^{-1}$, and is given by
 \begin{equation}
\langle s^{(+)} (\epsilon) \rangle_M = \frac{a k_C^2}{48 \pi^2}
\,.
 \end{equation}
Hence, the entropy computed in the volume $V_{\rm prop}$, by
considering that $T\sim a$ in the Unruh effect (see Eq.
(\ref{1})), is given by
 \begin{equation}
\langle S^{(+)}(\epsilon)\rangle_M \lesssim \frac{\cal A}{l_P^2}
\,.
 \end{equation}
Thus, also in the Rindler case the entropy is proportional to the
area of the event horizon (a result also obtained in Ref.
\cite{Laflamme:ec,Terashima:1999vw}), and bounded from above.

\noindent It is worth to note that in the case of Rindler
space-time, results are formally equivalent to the Schwarzschild
geometry since the surface gravity of the black hole is the
gravitational {\it acceleration} at radius $r$ measured at the
infinity.

 \end{document}